\documentclass[11pt]{article}
\usepackage{geometry}                
\usepackage{graphicx}
\usepackage{hyperref}
\usepackage{amssymb}
\usepackage{natbib}
\usepackage{authblk}
\title{Sashimi plots: Quantitative visualization of RNA sequencing read alignments}

\author[1,2,*]{Yarden Katz}
\affil[1]{Dept. of Brain and Cognitive Sciences, MIT, Cambridge, MA}
\affil[2]{Dept. of Biology, MIT, Cambridge, MA}
\affil[3]{The Broad Institute of MIT and Harvard, Cambridge, MA}
\affil[4]{Dept. of Biological Engineering, Cambridge, MA}
\affil[5]{Dept. of Statistics, Harvard University, Cambridge, MA}
\author[2,*]{Eric T. Wang}
\author[3,*]{Jacob Silterra} 
\author[3]{Schraga Schwartz} 
\author[3]{Bang Wong}
\author[3]{Jill P. Mesirov} 
\author[5,3]{Edoardo M. Airoldi}
\author[2,4]{Christopher B. Burge}
\affil[*]{These authors contributed equally.}

\begin{document}
\maketitle


\abstract{We introduce Sashimi plots, a quantitative multi-sample visualization of mRNA sequencing reads aligned to gene annotations. Sashimi plots are made using alignments (stored in the SAM/BAM format) and gene model annotations (in GFF format), which can be custom-made by the user or obtained from databases such as Ensembl or UCSC. We describe two implementations of Sashimi plots: (1) a stand-alone command line implementation aimed at making customizable publication quality figures, and (2) an implementation built into the Integrated Genome Viewer (IGV) browser, which enables rapid and dynamic creation of Sashimi plots for any genomic region of interest, suitable for exploratory analysis of alternatively spliced regions of the transcriptome. Isoform expression estimates outputted by the MISO program can be optionally plotted along with Sashimi plots. Sashimi plots can be used to quickly screen differentially spliced exons along genomic regions of interest and can be used in publication quality figures. The Sashimi plot software and documentation is available from: \url{http://genes.mit.edu/burgelab/miso/docs/sashimi.html}}

\section{Background}
Recent studies using high-throughput sequencing of mRNAs (RNA-Seq) revealed that most human genes produce alternatively spliced mRNAs, and that distinct mRNA isoforms can vary in expression profiles across tissues and organisms (\cite{WangETNature, BarbosaMoraisSplicing}), in a manner determined by a combination of protein \textit{trans} factors and cis-regulatory elements (\cite{WangSplicingCode}). In spite of the rapid increases in sequencing depth, determining the abundance level of distinct isoforms is statistically challenging, since the majority of reads sequenced are consistent with multiple isoforms (\cite{KatzMISO}). For this reason, it is also challenging to visualize the reads alongside the annotated gene isoforms in a way that allows for rapid comparisons of isoform expression across multiple samples or conditions to detect differential exon usage. To address these challenges, we developed Sashimi plot, a quantitative visualization of RNA-Seq reads alignments. Sashimi plots summarize the raw read alignments and are suitable for simultaneously displaying multiple RNA-Seq samples. 

\section{Sashimi plot visualizations}
\subsection{Features and inputs}
Sashimi plots are made using gene model annotations along with read alignments to generate a quantitative summary of the genomic and splice junction reads, shown schematically in Figure 1A. Genomic reads are converted into read densities (per base) scaled by the number of mapped reads in the sample, measured in RPKM units (\cite{MortazaviNatMethods}). Splice junction reads are plotted as arcs whose width is proportional to the number of junction reads that span the exons connected by the arc. In Figure 1A, reads across the two grey exons are used to the produce the Sashimi plot on right (in red).
Sashimi plots require two main inputs, shown in Figure 1B: (1) Alignments of reads to the genome (including junctions), provided in the standard BAM format. Read mappers that produce splice junction alignments, such as Tophat, can produce these. (2) Annotation of gene models or alternatively spliced events in GFF3 format (\cite{GFF3}). These annotations can be downloaded from databases such as Ensembl or UCSC, or custom-generated by the user (e.g. based on de novo transcript assembly programs.) We have generated a number of alternative isoform annotations in commonly studied genomes (available from the MISO website, \cite{MISOWebsite}) which can be used with Sashimi plots. A third optional input includes quantitative estimates of 
isoform abundance ($\Psi$ values), as estimated by MISO, which can be displayed alongside the Sashimi plots. 

The plots can be made using one of two implementations: a command-line Sashimi plot implementation that produces a static and highly customizable, publication-quality figures, or through the Broad IGV browser (``IGV-Sashimi'') (\cite{IGVPaper}), which enables dynamic on-the-fly creation of Sashimi plots for genomic regions of interest (Figure 1B). The static view Sashimi plot is configurable through an external settings file that allows users to customize the colors, scales, fonts and size of Sashimi plots. If given the output of the isoform quantitation program MISO, Sashimi plots can be automatically decorated with estimates of the abundance levels of alternative exons. 
The dynamic IGV-Sashimi is fully integrated with the genome browser and enables rapid zooming in and out of genomic regions of interest, which can then be rendered on-the-fly as Sashimi plots. IGV-Sashimi is aimed at rapid exploratory analyses of genomic regions, where differential usage of alternative splicing can be made apparent. Coverage criteria for reads and junctions can be set by the user, and the resulting Sashimi plot can be saved as an image or vector graphic.

\section{Interpreting Sashimi plots}
An example Sashimi plot across four RNA-Seq samples is shown in Figure 2A. Samples are color-coded by condition, with two RNA-Seq samples from wild type mice in red (`heartWT1', `heartWT2') and mouse heart tissues depleted for the splicing factor Muscleblind1 (`heartKOa', `heartKOb') in orange. Read densities across exons are quantified in RPKM units \cite{MortazaviNatMethods} and junction reads are plotted as arcs that are annotated with the raw number of junction reads present in each sample. The two alternative transcripts in the annotation (obtained from a user-provided GFF file) are shown at the bottom. The plot highlights the differential splicing of the middle exon, which is mostly included in the wild type heart samples, but mostly excluded in the knockout samples. This is confirmed by the MISO estimates for the expression of the exon (Figure 2A, right) where the `Percent Spliced In' ($\Psi$ value as in \cite{KatzMISO}) for the exon is $\sim$77$\%$ in the wild types samples and is reduced to $\sim$25$\%$ in the knockout samples.

An analogous plot can be generated using IGV-Sashimi. The genomic region containing the alternative exon is shown in the IGV browser in Figure 2B. A Sashimi plot generated from this region is (Figure 2B, on right), where the wild type heart sample is plotted in red and the knockout heart sample in blue. The annotation containing the alternatively spliced exon is shown at the bottom (read from a GFF file), with RefSeq canonical transcripts shown on top. The boundaries of the Sashimi plot are determined by the region of interest shown in the IGV browser window, and can be altered to include more or fewer exons using the zoom in/out feature of the browser. As in Figure 2A, the raw junction read counts are shown on top of each junction arc.

Both variants of Sashimi plot highlight that the alternative exon considered is differentially regulated between control and knockout sample groups. Differential exons of this sort can be rapidly screened by surveying regions in the IGV Browser, and rapidly quantified with IGV-Sashimi. When a genomic region of interest such as the alternative exon in Figure 2A is identified and selected for use in a publication-quality figure, it can be plotted in a customized way using the static Sashimi plot.

\subsection{Implementation}
The command-line version of Sashimi plot is written in Python and relies on the Python plotting library matplotlib (\cite{Matplotlib}). It is packaged as part of MISO and can be downloaded from {\tt pypi}, a central public repository of Python packages. IGV-Sashimi plot is available as part of the IGV browser implemented in Java. Both versions are compatible with Unix-based, Windows and Mac OS X operating systems. Both implementations are open source and available through publicly hosted Git repositories. 

\subsection{Manual and installation}
Sashimi plot manual and installation instructions are available at the Sashimi plot website (\cite{SashimiWebsite}).

\section{Conclusions and future work}
Sashimi plots allow quantitative and comparative visualization of RNA-Seq reads across different samples, with the aim of detecting differentially spliced exons and isoforms. Static Sashimi plots are suitable for publication quality figure generation, while IGV-Sashimi is geared toward rapid exploratory analyses of genomic regions which are not readily available in ordinary genome browsers. 

Several genome browser, such as The UCSC Genome Browser (\cite{UCSCGenomeBrowser}) can display read densities across exons (while showing isoforms in separate track), produced from BAM files or other formats. However, junction reads can only be displayed individually, with a line segment drawn for each read present in the sample, in the UCSC browser. This representation does not readily support quantitative comparisons among samples, and gets unwieldy for high-coverage samples where there are hundreds to thousands of junction reads. It would be useful to combine the multitude of useful annotation and data tracks of UCSC with Sashimi plots, so that differential exon usage can be overlapped with these tracks. 

Future work in the visualization domain would be to consider alternative representations of junction read counts. The current implementation of Sashimi plots uses arc width to represent counts, but other, potentially more visually quantitative forms could be considered (such as circle diameter/area, or a color shade-based representation.) We invite members of the community to build on, extend and modify Sashimi plots to capture more of the richness of information present in RNA sequencing data.

\section{Acknowledgements}

We thank James Robinson, Helga Thorvaldsd\'{o}ttir and Noam Shoresh of the Broad Institute for insightful discussion and comments on the manuscript.

\begin{figure}[t]
  \caption{{\bf Anatomy of a Sashimi plot and its inputs.} (A) Gene model annotation showing two transcripts (middle exon is alternatively spliced.) Sashimi plot for the two grey exons is shown, where genomic reads are converted into read densities (measured in RPKM) and junction reads are plotted as arcs whose width is determined by the number of reads aligned to the junction spanning the exons connected by the arc. (B) Inputs required for making a Sashimi plot. Gene model annotations (in GFF format), RNA-Seq read alignments (BAM format) and optionally isoform expression estimated (provided by the MISO program) are used to make a Sashimi plot. Sashimi plots for regions of interest can be made with a stand-alone program that makes customizable publication quality figures, or directly from the IGV browser.}
  \centering
    \includegraphics{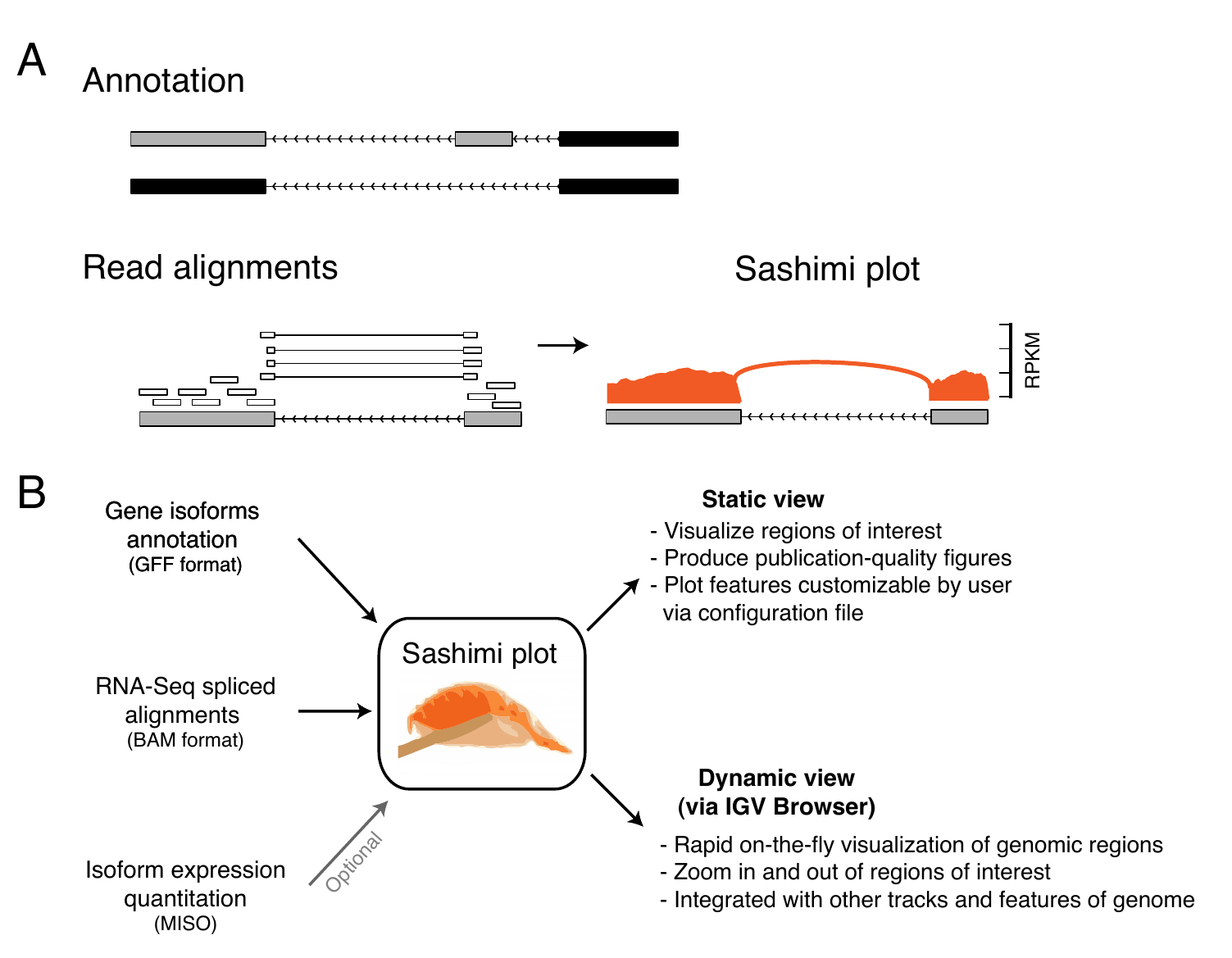}
\end{figure}

\begin{figure}[t]
  \caption{{\bf Example Sashimi plot for an alternatively spliced exon.}
(A) Gene model annotation showing two transcripts (middle exon is alternatively spliced.) Sashimi plot for the two grey exons is shown, where genomic reads are converted into read densities (measured in RPKM) and junction reads are plotted as arcs whose width is determined by the number of reads aligned to the junction spanning the exons connected by the arc. (B) Inputs required for making a Sashimi plot. Gene model annotations (in GFF format), RNA-Seq read alignments (BAM format) and optionally isoform expression estimated (provided by the MISO program) are used to make a Sashimi plot. Sashimi plots for regions of interest can be made with a stand-alone program that makes customizable publication quality figures, or directly from the IGV browser.
}
  \centering
    \includegraphics[scale=0.7]{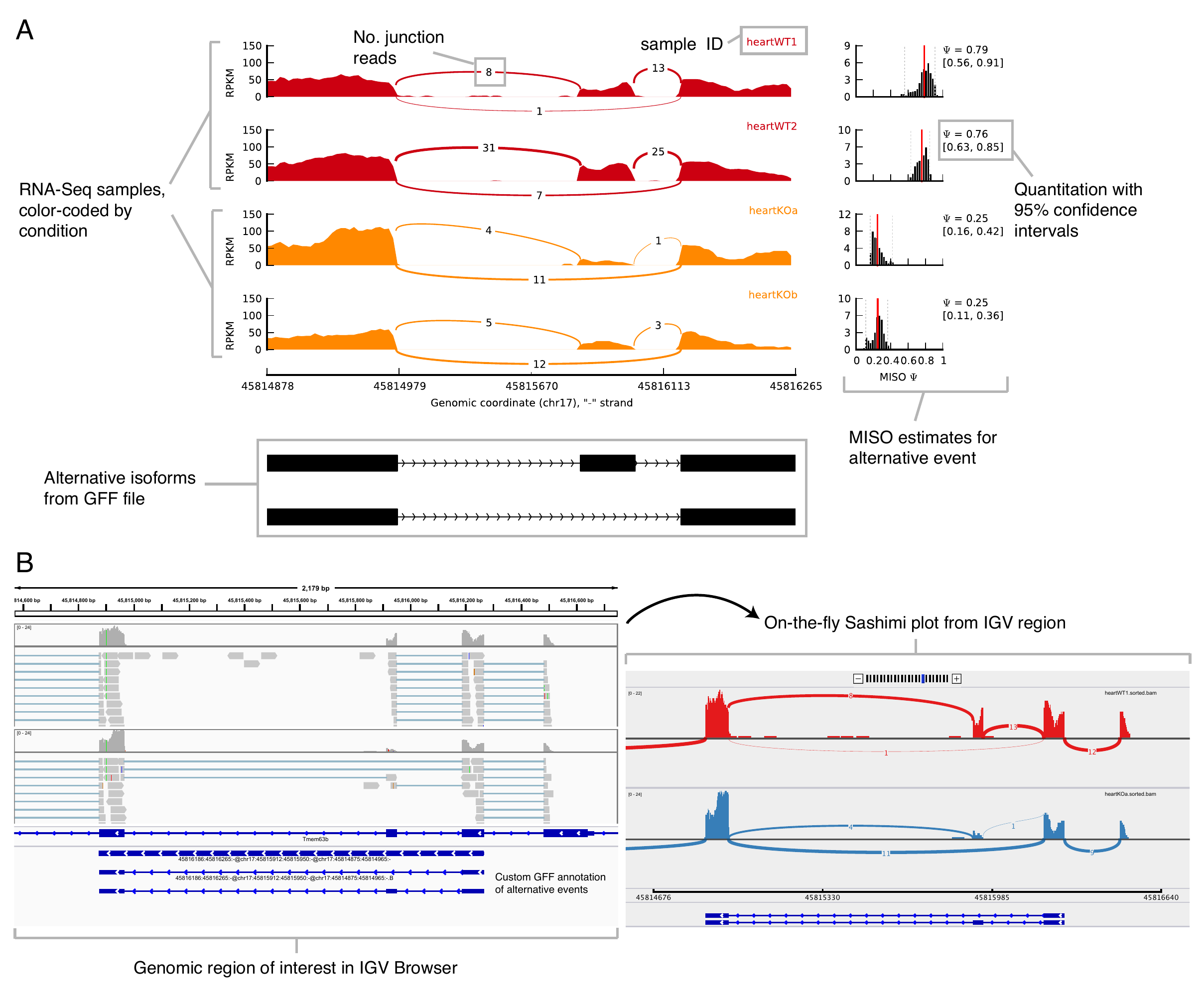}
\end{figure}

\bibliographystyle{plainnat}
\bibliography{sashimi_refs}

\begin{thebibliography}{11}
\providecommand{\natexlab}[1]{#1}
\providecommand{\url}[1]{\texttt{#1}}
\expandafter\ifx\csname urlstyle\endcsname\relax
  \providecommand{\doi}[1]{doi: #1}\else
  \providecommand{\doi}{doi: \begingroup \urlstyle{rm}\Url}\fi

\bibitem[GFF()]{GFF3}


\bibitem[Barbosa-Morais et~al.(2012)Barbosa-Morais, Irimia, Pan, Xiong,
  Gueroussov, Lee, Slobodeniuc, Kutter, Watt, Colak, Kim, Misquitta-Ali,
  Wilson, Kim, Odom, Frey, and Blencowe]{BarbosaMoraisSplicing}
N.~L. Barbosa-Morais, M.~Irimia, Q.~Pan, H.~Y. Xiong, S.~Gueroussov, L.~J. Lee,
  V.~Slobodeniuc, C.~Kutter, S.~Watt, R.~Colak, T.~Kim, C.~M. Misquitta-Ali,
  M.~D. Wilson, P.~M. Kim, D.~T. Odom, B.~J. Frey, and B.~J. Blencowe.
\newblock {{T}he evolutionary landscape of alternative splicing in vertebrate
  species}.
\newblock \emph{Science}, 338\penalty0 (6114):\penalty0 1587--1593, Dec 2012.

\bibitem[Hunter(2007)]{Matplotlib}
J.~D. Hunter.
\newblock Matplotlib: A 2d graphics environment.
\newblock \emph{Computing In Science \& Engineering}, 9\penalty0 (3):\penalty0
  90--95, 2007.

\bibitem[Katz et~al.(2010)Katz, Wang, Airoldi, and Burge]{KatzMISO}
Y.~Katz, E.~T. Wang, E.~M. Airoldi, and C.~B. Burge.
\newblock {{A}nalysis and design of {R}{N}{A} sequencing experiments for
  identifying isoform regulation}.
\newblock \emph{Nat. Methods}, 7\penalty0 (12):\penalty0 1009--1015, Dec 2010.

\bibitem[Kent et~al.(2002)Kent, Sugnet, Furey, Roskin, Pringle, Zahler, and
  Haussler]{UCSCGenomeBrowser}
W.~J. Kent, C.~W. Sugnet, T.~S. Furey, K.~M. Roskin, T.~H. Pringle, A.~M.
  Zahler, and D.~Haussler.
\newblock {{T}he human genome browser at {U}{C}{S}{C}}.
\newblock \emph{Genome Res.}, 12\penalty0 (6):\penalty0 996--1006, Jun 2002.

\bibitem[MISO()]{MISOWebsite}
MISO.
\newblock {MISO} website.
\newblock \url{http://genes.mit.edu/burgelab/miso/}.

\bibitem[Mortazavi et~al.(2008)Mortazavi, Williams, McCue, Schaeffer, and
  Wold]{MortazaviNatMethods}
A.~Mortazavi, B.~A. Williams, K.~McCue, L.~Schaeffer, and B.~Wold.
\newblock {{M}apping and quantifying mammalian transcriptomes by
  {R}{N}{A}-{S}eq}.
\newblock \emph{Nat. Methods}, 5\penalty0 (7):\penalty0 621--628, Jul 2008.

\bibitem[Sashimi()]{SashimiWebsite}
Sashimi.
\newblock Sashimi plot website.
\newblock \url{http://genes.mit.edu/burgelab/miso/docs/sashimi.html}.

\bibitem[Thorvaldsdottir et~al.(2013)Thorvaldsdottir, Robinson, and
  Mesirov]{IGVPaper}
H.~Thorvaldsdottir, J.~T. Robinson, and J.~P. Mesirov.
\newblock {{I}ntegrative {G}enomics {V}iewer ({I}{G}{V}): high-performance
  genomics data visualization and exploration}.
\newblock \emph{Brief. Bioinformatics}, 14\penalty0 (2):\penalty0 178--192, Mar
  2013.

\bibitem[Wang et~al.(2008)Wang, Sandberg, Luo, Khrebtukova, Zhang, Mayr,
  Kingsmore, Schroth, and Burge]{WangETNature}
E.~T. Wang, R.~Sandberg, S.~Luo, I.~Khrebtukova, L.~Zhang, C.~Mayr, S.~F.
  Kingsmore, G.~P. Schroth, and C.~B. Burge.
\newblock {{A}lternative isoform regulation in human tissue transcriptomes}.
\newblock \emph{Nature}, 456\penalty0 (7221):\penalty0 470--476, Nov 2008.

\bibitem[Wang and Burge(2008)]{WangSplicingCode}
Z.~Wang and C.~B. Burge.
\newblock {{S}plicing regulation: from a parts list of regulatory elements to
  an integrated splicing code}.
\newblock \emph{RNA}, 14\penalty0 (5):\penalty0 802--813, May 2008.

\end{thebibliography}
\end{document}